\documentclass[12pt, a4paper, reprint, aps, prd, nobibnotes, floatfix, showkeys, superscriptaddress]{revtex4-1}

\usepackage[utf8]{inputenc}
\usepackage{amsmath,amsthm,amssymb,amsfonts}
\usepackage{graphicx}
\usepackage{booktabs}
\usepackage{multirow}
\usepackage{xcolor}
\usepackage{rotating}
\usepackage[colorlinks]{hyperref}
\usepackage{orcidlink}
\usepackage{adjustbox}
\usepackage{array}
\usepackage{appendix}
\usepackage{units}
\usepackage[normalem]{ulem} 
\usepackage{comment}

\AtBeginDocument{
  \heavyrulewidth=.08em
  \lightrulewidth=.05em
  \cmidrulewidth=.03em
  \belowrulesep=.65ex
  \belowbottomsep=0pt
  \aboverulesep=.4ex
  \abovetopsep=0pt
  \cmidrulesep=\doublerulesep
  \cmidrulekern=.5em
  \defaultaddspace=.5em
}

\usepackage{array}
\newcommand{\PreserveBackslash}[1]{\let\temp=\\#1\let\\=\temp}
\newcolumntype{C}[1]{>{\PreserveBackslash\centering}p{#1}}

\newcommand{\asharp}[1]{$\text{A}\#_\text{3Sus}$}

\newcommand{\cbo}[2]{CE  $\unit[#1]{km}$ (CBO#2)}

\newcommand{\pmo}[2]{CE  $\unit[#1]{km}$ (PMO#2)}

\newcolumntype{R}[2]{
    >{\adjustbox{angle=#1,lap=\width-(#2)}\bgroup}
    l
    <{\egroup}
}

\begin{document}

\title{\texorpdfstring{Multi-messenger astronomy with a \\Southern-Hemisphere gravitational-wave observatory}{Multi-messenger astronomy with a Southern-Hemisphere gravitational-wave observatory}}

\author{James~W.~Gardner\,\orcidlink{0000-0002-8592-1452}}

\email{james.gardner@anu.edu.au}
\affiliation{OzGrav-ANU, Centre for Gravitational Astrophysics, Research Schools of Physics, and of Astronomy and Astrophysics, The Australian National University, Canberra ACT 2601, Australia}
\affiliation{Walter Burke Institute for Theoretical Physics, The California Institute of Technology, Pasadena, California 91125, USA} 
\author{Ling~Sun\,\orcidlink{0000-0001-7959-892X}}
\affiliation{OzGrav-ANU, Centre for Gravitational Astrophysics, Research Schools of Physics, and of Astronomy and Astrophysics, The Australian National University, Canberra ACT 2601, Australia}
\author{Ssohrab~Borhanian\,\orcidlink{0000-0003-0161-6109}}
\affiliation{Theoretisch-Physikalisches Institut, Friedrich-Schiller-Universit\"at Jena, 07743, Jena, Germany}
\author{Paul~D.~Lasky\,\orcidlink{0000-0003-3763-1386}}
\author{Eric~Thrane\,\orcidlink{0000-0002-4418-3895}}
\affiliation{School of Physics and Astronomy, Monash University, VIC 3800, Australia}
\affiliation{OzGrav: The ARC Centre of Excellence for Gravitational Wave Discovery, Clayton VIC 3800, Australia}
\author{David~E.~McClelland\,\orcidlink{0000-0001-6210-5842}}
\author{Bram~J.~J.~Slagmolen\,\orcidlink{0000-0002-2471-3828}\,}
\affiliation{OzGrav-ANU, Centre for Gravitational Astrophysics, Research Schools of Physics, and of Astronomy and Astrophysics, The Australian National University, Canberra ACT 2601, Australia}

\date{\today}

\begin{abstract}
Joint observations of gravitational waves and electromagnetic counterparts will answer questions about cosmology, gamma-ray bursts, and the behaviour of matter at supranuclear densities.
The addition of a Southern-Hemisphere gravitational-wave observatory to proposed global networks creates a longer baseline, which is beneficial for sky localisation.
We analyse how an observatory in Australia can enhance the multi-messenger astronomy capabilities of future networks.
We estimate the number of binary neutron star mergers with joint observations of gravitational waves and kilonova counterparts detectable by the Vera C.\ Rubin Observatory. 
First, we consider a network of upgrades to current observatories.
Adding an Australian observatory to a three-observatory network (comprising two observatories in the USA and one in Europe) boosts the rate of joint observations from $\unit[2.5^{+4.5}_{-2.0}]{yr^{-1}}$ to $\unit[5.6^{+10}_{-4.5}]{yr^{-1}}$ (a factor of two improvement).
Then, we consider a network of next-generation observatories.
Adding a $\unit[20]{km}$ Australian observatory to a global network of a Cosmic Explorer $\unit[40]{km}$ in the USA and an Einstein Telescope in Europe only marginally increases the rate from $\unit[40^{+71}_{-32}]{yr^{-1}}$ to $\unit[44^{+79}_{-35}]{yr^{-1}}$ (a factor of 1.1 improvement). 
The addition of an Australian observatory, however, ensures that at least two observatories are online far more often. 
When the Cosmic Explorer $\unit[40]{km}$ is offline for a major upgrade, the Australian observatory increases the joint observation rate from $\unit[0.5^{+0.8}_{-0.4}]{yr^{-1}}$ to $\unit[38^{+68}_{-30}]{yr^{-1}}$ (a factor of 82 improvement). 
When the Einstein Telescope is offline, the joint observation rate increases from $\unit[0.2^{+0.3}_{-0.1}]{yr^{-1}}$ to $\unit[19^{+34}_{-15}]{yr^{-1}}$ (a factor of 113 improvement).
We sketch out the broader science case for a Southern-Hemisphere gravitational-wave observatory.
\end{abstract}
\maketitle
\allowdisplaybreaks

\section{Introduction}
\label{sec:introduction}

Approximately 90 gravitational-wave signals have been observed to date from the mergers of black holes and neutron stars~\cite{GWTC-1:2018,GWTC-2:2020,GWTC-3:2021}. These signals have led to many scientific discoveries in cosmology, astrophysics, nuclear physics, and astronomy, but much remains to be discovered in the decades ahead~\cite{vitale2021first,cai_2017,miller2019new,Mancarella:2022GT}. The contributions of the two Laser Interferometric Gravitational-wave Observatories (LIGOs) in the USA~\cite{AdvancedLIGO:2015} and Virgo in Europe~\cite{acernese2014advanced}, were essential to the first detection of a binary neutron star merger~\cite{GW170817}. The joint third observing run of the global network also included the Kamioka Gravitational-Wave Observatory (KAGRA) in Japan~\cite{akutsu2018kagra,abbott2020prospects}. Looking to the future of the ground-based global network, one additional LIGO observatory is under construction in India~\cite{saleem2021science} and many proposals have been made for upgrades to the existing observatories~\cite{Asharp,LIGO_Voyager,Universe7090322,VirgoPostO5,10.1093/ptep/ptaa120}. Next-generation observatories, e.g., Cosmic Explorer (CE)~\cite{CosmicExplorerHorizonStudy2021} and the Einstein Telescope (ET)~\cite{EinsteinTelescope}, are also under development.

The success to date of gravitational-wave astronomy has been achieved through international collaboration and the burgeoning global network of observatories. 
The addition of a Southern-Hemisphere gravitational-wave observatory could provide a long baseline from the Northern-Hemisphere observatories to improve sky localisation. 
Several previous studies have proposed Australia as a potential host for a Southern-Hemisphere observatory~\cite{DGBlair_2008,NEMO_2020,sarin_lasky_2022,CosmicExplorerHorizonStudy2021}.

We consider how an Australian observatory can enhance the multi-messenger astronomy capabilities of the global network. 
Neutron stars are a promising tool for multi-messenger astronomy with gravitational waves.
They host one of the most extreme environments in the known Universe, far exceeding densities that can be studied on Earth. Observing gravitational waves from the merger of binary neutron stars offers a momentary glimpse into the stars' otherwise inaccessible, potentially exotic cores~\cite{lasky_2015,BAIOTTI2019103714,bauswein2019identifying,Universe7040097}. Two binary neutron star mergers, GW170817~\cite{GW170817} and GW190425~\cite{GW190425}, have been observed by the current generation of gravitational-wave observatories. 
The event GW170817 was constrained to a sky-area of $\unit[28]{deg^2}$ at a luminosity distance of $\unit[40^{+8}_{-14}]{Mpc}$ by Advanced LIGO and Virgo (Bayesian 90\%-credible intervals). A gamma-ray burst was observed by the Fermi Gamma-ray Burst Monitor about $\unit[1.7]{s}$ after the merger~\cite{abbott2017gravitational}. A subsequent follow-up campaign discovered a bright optical transient in the host galaxy NGC~4993~\cite{swopeGW170817}, which was identified as a kilonova, an electromagnetic transient powered by the radioactive decay of heavy nuclei produced in the ejecta, with shock fronts visible in X-ray and radio, e.g., see Refs.~\cite{GW170817multi,australianGW170817}. 

The joint gravitational and electromagnetic observation of GW170817 provided invaluable information about astrophysics, dense matter, and cosmology. Combining the distance of the source measured from the gravitational-wave signal and the recession velocity inferred from the redshift measurements from electromagnetic observations, GW170817 was used as a ``standard siren''~\cite{Holz2005} to provide an independent measure of the Hubble constant~\cite{GW170817H0}, later refined using radio observations of the jet~\cite{hotokezaka2019hubble,mooley2018superluminal}. Population studies of multi-messenger observations of binary neutron stars promise a variety of exciting astronomical and astrophysical discoveries including resolving the Hubble-Lema\^{i}tre tension~\cite{Pian+2021,NEMO_2020,sarin_lasky_2022,martynov2019exploring}. According to Ref.~\cite{Chen2018}, $\mathcal{O}(100)$ joint observations will lead to an approximately $1\%$ determination of the Hubble constant. 
The promise of multi-messenger astronomy from $\mathcal{O}(100)$ observations, however, is far broader than just cosmology. For example, it will provide unprecedented insight into the physics and astrophysics of gamma-ray bursts including jet launching and propagation, the formation mechanisms of neutron stars and neutron-star binaries, as well as precision studies on the hot and cold equations of state of matter at supranuclear densities; see, e.g., Refs.~\cite{10.1088/2514-3433/ac2256ch9,sarin_lasky_2022} and the references therein.

The rest of this paper is organised as follows.
In Section~\ref{sec:AustralianGWDs}, we describe different scenarios where an Australian observatory is added to the global network.
Then, in Section~\ref{sec:method}, we describe our method for estimating the improvement in multi-messenger capabilities due to adding the Australian observatory.
Finally, we present our results in Section~\ref{sec:results} and conclusions in Section~\ref{sec:conclusions}.

\section{Proposed gravitational-wave observatories in Australia}
\label{sec:AustralianGWDs}

\begin{figure}[t]
    \centering
    \includegraphics[width=\columnwidth]{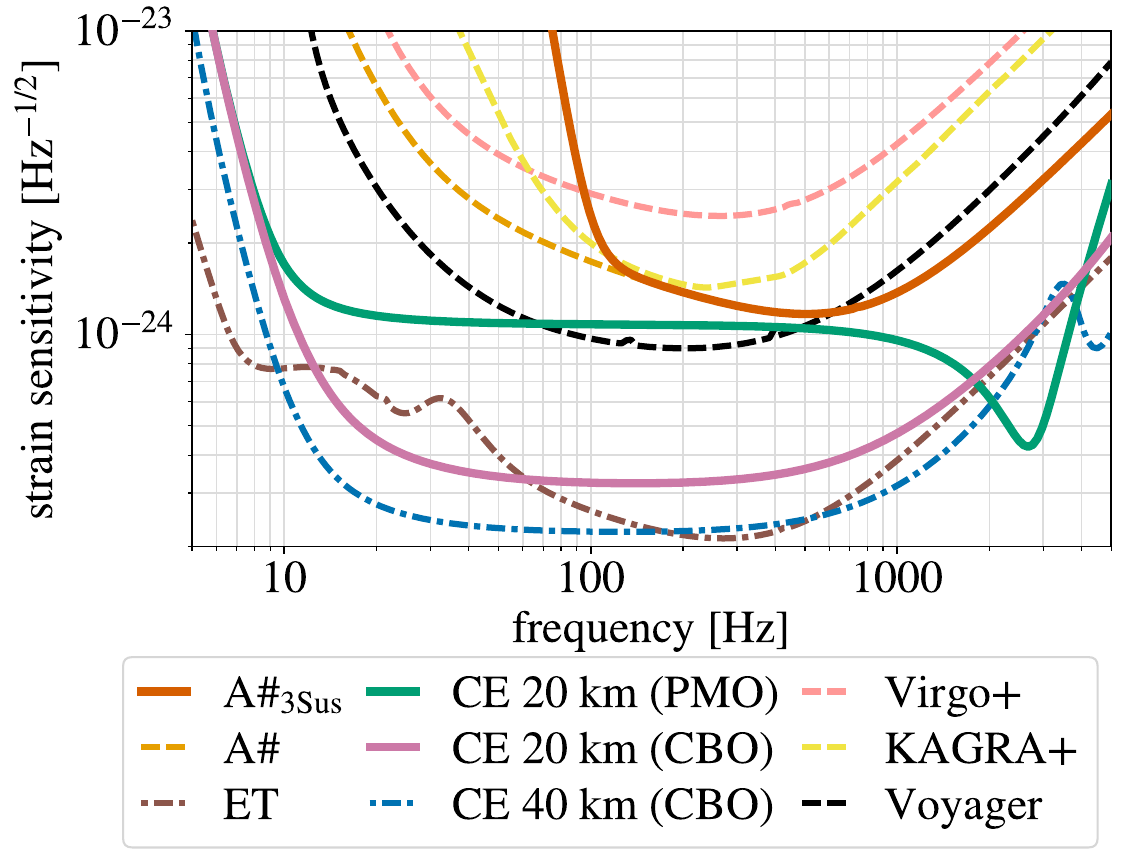}   
    \caption{Strain sensitivity versus frequency for proposed Australian gravitational-wave observatories considered in this work (solid curves) compared to the global context (dashed curves for the upgraded 2G--era and dashed-dotted curves for the XG-era)~\cite{Borhanian_2021}. Lower values of the strain sensitivity indicate greater sensitivity to gravitational waves. For future observatories, the labels refer only to the displayed sensitivity, not a particular technology used to achieve that sensitivity, e.g., for the compact-binary coalescence optimised (CBO) and post-merger optimised (PMO) designs for CE.
    }
    \label{fig:sensitivityCurves}
\end{figure}

We consider three potential Australian gravitational-wave observatories with sensitivities shown in Fig.~\ref{fig:sensitivityCurves}. 
We assume that each of these observatories is situated at 26°S and 116°E~\cite{Note1} with the Y-arm of the interferometer at due Northeast; we label this location and orientation as ``AU''. 
While we are primarily interested in the possibility of an Australian observatory, the results would be similar using another Southern-Hemisphere location with a similar baseline to the Northern-Hemisphere observatories. 

We do not consider all the possible Australian observatories proposed in the literature. 
The Neutron-star Extreme Matter Observatory (NEMO)~\cite{NEMO_2020,NEMOcurve,sarin_lasky_2022} is another proposal for an Australian observatory, however, its science case focuses on $\unit[\text{1--4}]{kHz}$ gravitational-wave physics while we focus on multi-messenger astronomy from broadband frequencies.
Further discussion about NEMO can be found in Appendix~\ref{app:NEMO}.

\subsection{Modified A\#}
\label{sec:Asharp}
The LIGO A\# concept~\cite{Asharp,LIGOdovetail} is a proposed upgrade to the present LIGO's dual-recycled Fabry-P\'{e}rot Michelson interferometers. 
The main upgrades for A\# are $\unit[100]{kg}$ test masses, improved seismic isolation and suspension systems, increased laser input power and quantum squeezing, and improved coating thermal noise. 
We consider an A\# observatory in Australia, which has been modified to reduce the financial cost and mitigate operational complexities for a potentially shorter commissioning time.
In particular, this modified version of A\# employs the original LIGO A+ test-mass coatings and a triple suspension system for the test-masses~\cite{Note7}, which reduces the sensitivity level below $\unit[100]{Hz}$ relative to the unmodified A\#. 
We refer to this modified, triple-suspension configuration as \asharp{}.

\subsection{Cosmic Explorer $\unit[20]{km}$}
\label{sec:CE}

We also consider a next-generation Cosmic Explorer (CE) observatory with $\unit[20]{km}$ arms in Australia~\cite{CosmicExplorerHorizonStudy2021,evans2023cosmic}. 
The CEs will be new $\unit[20]{km}$-long or $\unit[40]{km}$-long facilities at new locations. The observatories will utilise larger test masses, improved suspension and isolation systems, increased laser input power and quantum squeezing, and lower loss mirror coatings over the current generation of observatories.
There are two possible CE designs for the $\unit[20]{km}$ facility that we consider for the Australian observatory~\cite{CosmicExplorerHorizonStudy2021,evans2023cosmic,Srivastava2022.APJ}. The compact-binary coalescence optimised (CBO) configuration is designed for broadband sensitivity and, therefore, is particularly relevant to our science case. For completeness, we also study the post-merger optimised (PMO) configuration designed for enhanced kilohertz sensitivity. We refer to these configurations as \cbo{20}{} and \pmo{20}{}, respectively.

\subsection{Global network scenarios}\label{scenarios}
We define two eras or ``generations'' of the global gravitational-wave observatory network to study: future upgrades to current observatories referred to as ``upgraded 2G'', and new facilities operating with an order of magnitude better strain sensitivity referred to as ``XG''. 
We envision an upgraded 2G--era in which LIGO is upgraded to the Voyager design~\cite{LIGO_Voyager} which, amongst many changes, uses a different laser wavelength and cryogenic silicon test masses, alongside the Advanced Virgo+ (which we label as Virgo+)~\cite{Universe7090322} and KAGRA+~\cite{10.1093/ptep/ptaa120} designs. 
For the XG-era, we consider an Australian CE operating simultaneously with a CE $\unit[40]{km}$ in the USA~\cite{CosmicExplorerHorizonStudy2021}.
We also examine how an Australian CE would benefit other next-generation global networks including those involving ET~\cite{EinsteinTelescope}. 
We list various network scenarios in Table~\ref{tab:networkScenarios}.
Previous network studies have examined similar but not identical scenarios~\cite{gupta2023characterizing,Note8}. For future observatories, we assume fiducial locations for the purpose of calculation as the actual sites are not yet known.

\subsection{Observational duty cycle}
\label{sec:duty_cycle}
To fully assess the benefit of an Australian observatory, it is necessary to consider the network duty cycle --- the fraction of time for which the full network is observing.
During the third Advanced LIGO observing run, each LIGO achieved an individual duty cycle of 76--79\%~\cite{GWTC-3:2021,Davis}.
The ``coincident duty cycle'' when both LIGOs were operating simultaneously in a two-observatory network was approximately 58--62\% (assuming that the downtimes of different observatories are uncorrelated).
There are a number of reasons that gravitational-wave observatories do not continuously operate during an observing run, including regular maintenance and lock loss from environmental disturbances~\cite{Davis}.
These values for the duty cycle, however, do not account for the fact that gravitational-wave observatories regularly undergo long periods of downtime between observing runs for major upgrades and commissioning.
Accounting for this downtime between observing runs, the overall effective coincident duty cycle since the construction of Advanced LIGO is approximately 27\%.
It is important that at least two observatories are operating simultaneously to achieve the sky-localisation necessary for electromagnetic follow-up of binary neutron star mergers. For future networks particularly in the XG-era, therefore, we emphasise how a third observatory can be used to maximise the coincident duty cycle when two or more observatories are online.

\begin{table*}[ht]
    \centering
    \addtolength{\tabcolsep}{5pt} 
    \begin{tabular}{@{}rr|l@{}}
    \toprule
    generation & Australian (AU) observatory & global network 
    of observatories                    \\ \midrule
    upgraded 2G       & \asharp{}     & Voyager (H), Voyager (L)                        \\
               & \asharp{}     & Voyager (H), Voyager (L), Virgo+                  \\
               & \asharp{}     & Voyager (H), Voyager (L), Virgo+, KAGRA+            \\
               & \asharp{}     & Voyager (H), Voyager (L), Virgo+, KAGRA+, Voyager (I) \\ \midrule
    XG         & \cbo{20}{}        & \cbo{40}{; C}                                        \\
               & \cbo{20}{}        & ET                                        \\
               & \cbo{20}{}        & \cbo{40}{; C}, \cbo{20}{; N}                    \\
               & \cbo{20}{}        & ET, \cbo{40}{; C}                           \\[7pt] 
               & \pmo{20}{}        & \cbo{40}{; C}                                        \\
               & \pmo{20}{}        & ET                                        \\
               & \pmo{20}{}        & \cbo{40}{; C}, \cbo{20}{; N}                    \\
               & \pmo{20}{}        & ET, \cbo{40}{; C}                           \\ \bottomrule
    \end{tabular}
    \addtolength{\tabcolsep}{-5pt} 
    \caption{Network scenarios showing the proposed Australian observatories and a selection of possible global environments. 
    Cosmic Explorer (CE) has a variety of possible configurations including those labelled as ``CE $\unit[20]{km}$'' (``CE $\unit[40]{km}$'') for CE with $\unit[20]{km}$ ($\unit[40]{km}$) arms as well as the choice between the compact-binary coalescence optimised (CBO) and post-merger optimised (PMO) designs.
    We abbreviate the following locations: H for Hanford, USA; L for Livingston, USA; I for Hingoli, India; C for Idaho, USA; and N for New Mexico, USA. Virgo and KAGRA are modelled as being at their present sites. The Einstein Telescope (ET) is assumed to be in Cascina, Italy.
    }
    \label{tab:networkScenarios}
\end{table*}

\section{Method}
\label{sec:method}

We benchmark each network scenario by simulating a universe of sources, predicting how the network observes each source, summarising the total performance over the population, and addressing known biases. We detail each of these steps in turn.

\subsection{Monte Carlo population}
\label{sec:population}

We focus on optical and ultraviolet observations of kilonovae emissions and use the Vera C.\ Rubin Observatory (Rubin)~\cite{thomas2020vera} as our standard. Rubin's typical survey depth for a GW170817-like kilonova given an exposure time of $\unit[30]{s}$ is $\unit[500]{Mpc}$ (i.e., redshift $\approx0.1$). We call sources ``in range'' within this distance henceforth. Within range, we can still observe other counterparts, e.g., gamma-rays, X-rays, and radio waves. We defer a complete study of the capabilities of multi-messenger astronomy beyond $\unit[500]{Mpc}$ to future work.

We simulate a population of binary neutron star mergers in range. We calculate the redshift distribution of sources in Appendix~\ref{app:merger_rate}. The \mbox{GWTC-3} local merger rate of 
$\unit[106^{+190}_{-84}]{yr^{-1}Gpc^{-3}}$~\cite{GWTC-3:2021} implies $\unit[44^{+79}_{-35}]{yr^{-1}}$ sources in range although we simulate a larger, i.e., multi-year, population. The merger rate is only loosely constrained since the width of the 90\%-credible interval (the uncertainty range quoted above) is greater than twice the maximum a~posteriori estimate obtained from the Bayesian parameter estimation. This is the largest uncertainty in our analysis. In comparison, we estimate that the sampling error from the multi-year Monte Carlo population contributes an uncertainty of only $\mathcal{O}(1)$ source per year in the observation rate. 
To mitigate the uncertainty in the merger rate, since it is ultimately only a normalisation factor, we study the observation rate normalised to the total number of sources in range per year.

For the other parameters of the population, the right ascension, sine of the declination, polarisation angle, and cosine of the inclination angle all follow uniform distributions. The chirp mass, mass ratio, and angular momenta follow a cosmological model detailed in Ref.~\cite{BS2022}. We use the \texttt{IMRPhenomD\_NRTidalv2} waveform (which is an aligned-spin model) from the LIGO Scientific Collaboration Algorithm Library Suite (\textsc{LALSuite})~\cite{lalsuite} to model the binary neutron star merger signals.

\subsection{Fisher information analysis}
\label{sec:fisher}

To determine how well a network observes a given source, we use the Fisher information analysis tool \mbox{\textsc{GWBENCH}}~\cite{Borhanian_2021} written in \textsc{Python}~\cite{python} which is similar to other Fisher benchmarking tools~\cite{gwfish,gwfast}. We calculate the Cram\'er-Rao bound on the parameter estimates, specifically the sky-area, by computing the gradient of the measured gravitational-wave strain with respect to each astrophysical parameter.
In the single-parameter case~\cite{Note3}, the Cram\'er-Rao bound provides a lower bound on the mean-square-error $\text{E}[(\hat \theta - \theta)^2]$ for an estimate $\hat\theta$ of a parameter $\theta$. For example, when estimating the true mean $\theta$ of a Gaussian random variable $X\sim \text{N}(\theta,\sigma)$, the Cram\'er-Rao bound of $\sigma^2/N$ for $N$ observations is achieved by the sample mean $\hat\theta$. 
We provide technical specifications of our computation in Appendix~\ref{app:computation}. We recover the signal-to-noise ratio and sky-area estimates from Ref.~\cite{BS2022} which use the same method~\cite{Note2}.

Compared to Bayesian analysis, the Fisher information approach is computationally less expensive but requires careful handling of rejection errors discussed later~\cite{rodriguez2013inadequacies}. The reliability of Fisher analysis and specifically the bound on the sky-area compared to Bayesian analysis improves with the network signal-to-noise ratio~\cite{vallisneri2008use,Magee_2022,Iacovelli_2022}. For non-Gaussian distributions, the Cram\'er-Rao lower bound can differ significantly from the achievable minimal error. For example, at low signal-to-noise ratios $\lesssim 10$, Fisher analysis systematically underestimates the sky-area. While at signal-to-noise ratios $\gtrsim 25$, Ref.~\cite{Magee_2022} shows that the Fisher and Bayesian sky-area estimates roughly agree at the population level. Although there remains some variance in the individual estimates, there is little systematic bias between the two approaches. In particular, the number of sources with signal-to-noise ratio $\gtrsim 25$ that are localised to within $\unit[10]{deg^2}$ is consistent.

\subsection{Follow-up metrics for benchmarking}
\label{sec:metrics}

We study sources satisfying the following conditions:
\begingroup
\allowdisplaybreaks
\begin{enumerate}
    \item \emph{A signal-to-noise ratio threshold.} --- For the Bayesian and Fisher estimates to agree at the population level, we require the network signal-to-noise ratio to be greater than 25. We also examine other 2G-era networks with \asharp{} alongside the LIGOs operating at the A\# rather than the Voyager sensitivity. For these networks, however, less than 5\% of sources in range have a signal-to-noise ratio greater than 25 even for \asharp{} with five Northern Hemisphere observatories. We omit these networks from the results. 
    \item \emph{A luminosity distance threshold.} --- To observe a source with electromagnetic telescopes, it needs to be within their survey depth. We use Rubin's survey depth of $\unit[500]{Mpc}$ as discussed previously. For this threshold, we use the true luminosity distance to the source from our simulation, although this is not exactly known in practice.
    \item \emph{A sky localisation threshold.} --- To point the electromagnetic telescopes towards the source, the gravitational-wave observation has to sufficiently localise the source as quantified by the $90\%$-credible sky-area in square degrees. For context, GW170817 was localised to 
    $\unit[28]{deg^2}$~\cite{GW170817} and Rubin's field-of-view is $\unit[10]{deg^2}$~\cite{thomas2020vera,Note4}. Tighter localisation allows for faster and more precise identification of the host galaxy and potentially better multi-messenger astronomy. We model Earth's rotation as this assists in localisation for observatories with sufficient low-frequency sensitivity, e.g., \cbo{20}{}.
    For the sake of choosing a threshold for a simple counting metric, we examine sources localised ``loosely'' (within $\unit[10]{deg^2}$) and ``tightly'' (within $\unit[1]{deg^2}$). 
\end{enumerate}
\endgroup

As summarised in Table~\ref{tab:metrics}, we use the number of joint observations per year that satisfy the above three thresholds as our loose and tight follow-up metrics to benchmark different networks. As discussed previously, we normalise the metrics to the number of sources in range to define the ``fractional metrics''.

\addtolength{\tabcolsep}{5pt}    
\begin{table}[ht]
\centering
\begin{tabular}{@{}rll@{}}
\toprule
follow-up metric [$\text{yr}^{-1}$]           & loose          & tight        \\ \midrule
signal-to-noise ratio & $\geq25$ & $\geq25$ \\
luminosity distance [Mpc] & $\leq500$ & $\leq500$ \\
90\%-credible sky-area [$\text{deg}^2$] & $\leq10$       & $\leq1$      \\
 \bottomrule
\end{tabular}
\caption{Thresholds for the loose and tight follow-up metrics used to benchmark the networks for enabling electromagnetic follow-up of binary neutron star mergers.}
\label{tab:metrics}
\end{table}
\addtolength{\tabcolsep}{-5pt}

These metrics assume no improvements in the survey depth and localisation capabilities of electromagnetic telescopes in the coming decades, e.g., see Refs.~\cite{Bellm_2019,Ivezic_2019,sarin_lasky_2022}. These advancements would improve the metrics, particularly for XG. Future work should determine the likely impact of these improvements.

\subsection{Numerical rejections}

Our Fisher information analysis requires careful handling of numerically ill-conditioned matrices to avoid biasing the metrics. Ill-conditioned matrices, e.g., from nearly edge-on sources, are numerically inaccurate to invert. These sources are rejected from the analysis and not included in the metrics. The rejection rate is higher with fewer observatories in the network. We detail our approach to handling these rejections in Appendix~\ref{app:rejections}. To summarise, the fractional follow-up metrics are overestimated by an amount depending on the rejection scenario. The most realistic scenario lies somewhere between the worst-case scenario where the fractional metrics are uniformly reduced by a multiplicative factor of 0.32 (e.g., from 50\% to 16\%) and the ``null-case'' scenario where they remain unchanged.

To avoid this ambiguity, we provide a secondary figure-of-merit that is invariant of the merger and rejection rates.
We consider the geometric improvement in the follow-up metrics from adding an Australian observatory to a baseline network, i.e., the multiplicative factor of improvement in the number of joint observations. Whether or not we normalise the rates to the number of sources in range for Rubin does not affect this geometric improvement since the same normalisation factor appears in both the baseline network's rate and the network with an Australian observatory's rate and thus cancels out. Similarly, any factor from the merger rate or unified rejection rate (i.e., the rejection rate made equal among networks) cancels out. When the baseline network's metric is zero, the geometric improvement is undefined. In such cases, we provide a lower bound derived from the Monte Carlo resolution of our simulated population~\cite{Note5}. We emphasise that the geometric improvement is different from the arithmetic increase in the number of observations since the baseline network's performance is factored in when evaluating the geometric improvement. This makes the geometric improvement sensitive to small but nonzero values of metric from the baseline network. To understand the complete picture, we study both the arithmetic and geometric improvements in the number of joint observations.

\section{Results}
\label{sec:results}
\begin{figure*}[ht]
    \centering
    \includegraphics[width=\textwidth]{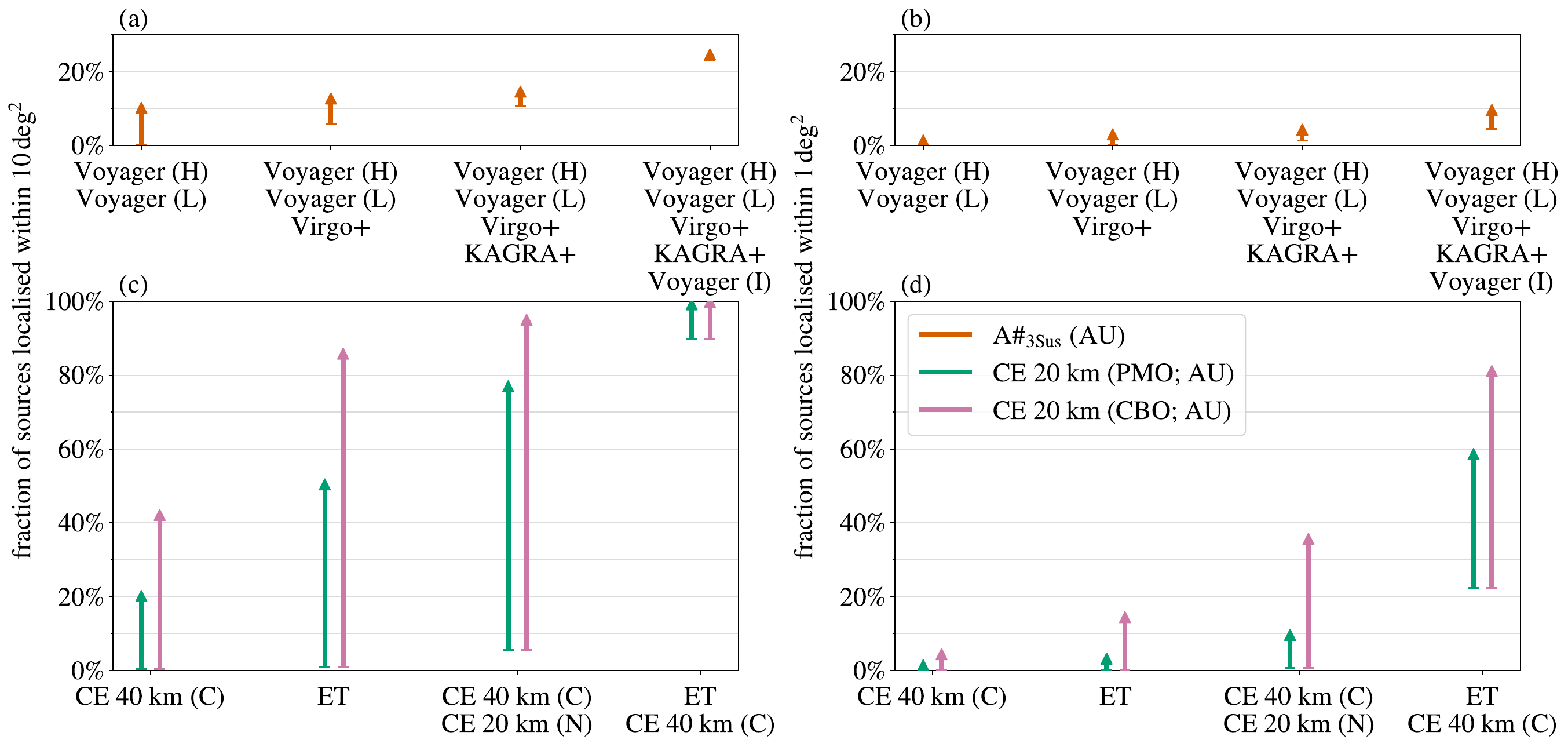}
    \caption{
    Fractional follow-up metrics from Table~\ref{tab:metrics}, i.e., the fraction of sources within $\unit[500]{Mpc}$ that have a network signal-to-noise ratio greater than 25 and are localised within $\unit[10]{deg^2}$ [panels (a) and (c)] or $\unit[1]{deg^2}$ [panels (b) and (d)].
    The length of each line shows the arithmetic improvement in the expected number of joint observations of gravitational waves and electromagnetic counterparts (normalised to the number of sources in range) from adding an Australian observatory to the baseline global networks in Table~\ref{tab:networkScenarios} in the upgraded 2G--era [panels (a) and (b)] and XG-era [panels (c) and (d)]. 
    We assume the compact-binary coalescence optimised (CBO) design for the Northern Hemisphere CEs but study the CBO and post-merger optimised (PMO) designs for the Australian CE.
    }
    \label{fig:metric}
\end{figure*}

\begin{figure*}[ht]
    \centering
    \includegraphics[width=\textwidth]{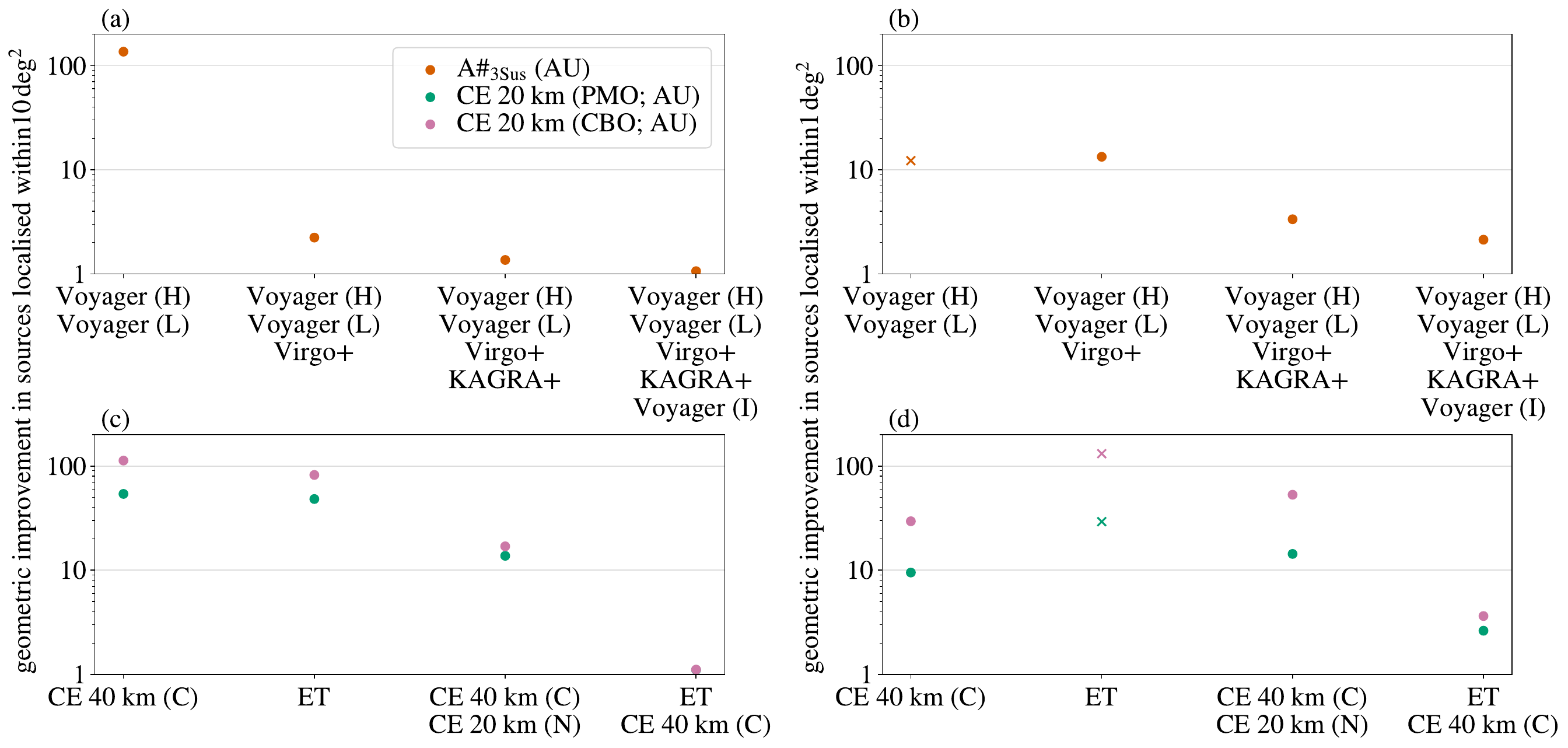}
    \caption{
    Geometric improvement in the follow-up metrics from Table~\ref{tab:metrics}, i.e., the factor of improvement in the number of observations within $\unit[500]{Mpc}$ that have a network signal-to-noise ratio greater than 25 and are localised within $\unit[10]{deg^2}$ [panels (a) and (c)] or $\unit[1]{deg^2}$ [panels (b) and (d)].
    This shows the multiplicative increase --- compared to the arithmetic increase shown in Fig.~\ref{fig:metric} --- in the expected number of joint observations of gravitational waves and electromagnetic counterparts from adding an Australian observatory to the baseline global networks in Table~\ref{tab:networkScenarios} in the upgraded 2G--era [panels (a) and (b)] and XG-era [panels (c) and (d)].
    We assume the compact-binary coalescence optimised (CBO) design for the Northern Hemisphere CEs, but study the CBO and post-merger optimised (PMO) designs for the Australian CE.
    When the baseline network observes zero sources, we provide a lower bound on the geometric improvement indicated by $\times$.
    }
    \label{fig:lambda}
\end{figure*}

We examine how a long-baseline global network with an observatory in Australia improves the joint observation rate quantified by the fractional follow-up metrics.

\subsection{Upgraded 2\textsuperscript{nd} generation (upgraded 2G)}

Fig.~\hyperref[fig:metric]{2a} shows the arithmetic improvement in the fractional loose follow-up metric when \asharp{} (AU) is added to each upgraded 2G--era global environment. For example, with three Northern Hemisphere observatories (Voyager (H), Voyager (L), and Virgo+) we observe 6\% of the sources within $\unit[500]{Mpc}$ that are localised within $\unit[10]{deg^2}$ and have a network signal-to-noise ratio above 25. Adding \asharp{} (AU) to this network increases the fraction to 13\% of the sources in range.
This improvement comes mainly from better sky localisation rather than from increasing the signal-to-noise ratio.
Fig.~\hyperref[fig:lambda]{3a} shows the geometric improvement of $2.2$ which is invariant of the unified rejection rate.

Overall, Fig.~\hyperref[fig:lambda]{3a} shows that \asharp{} (AU) improves the fractional loose follow-up metric by a multiplicative factor of at least 1.1 across all upgraded 2G--era scenarios. The greatest improvement seen in Fig.~\hyperref[fig:metric]{2a} is for the network with two Northern Hemisphere observatories (Voyager (H) and Voyager (L)) where adding \asharp{} (AU) improves the metric from approximately 0\% to 10\% of the sources in range.
With \asharp{} (AU), Voyager (H), and Voyager (L), this corresponds to observing $\unit[4.5^{+8.1}_{-3.6}]{yr^{-1}}$ sources accounting for the uncertainty in the GWTC-3 merger rate which dominates the other uncertainties in our method such as the rejection scenario and the Monte Carlo error. This rate increases to $\unit[11^{+20}_{-9}]{yr^{-1}}$ with \asharp{} (AU) and five Northern Hemisphere observatories such that it would take $\mathcal{O}(9)$ years to observe $\mathcal{O}(100)$ sources.

The upgraded 2G--era results for the tight metric from Table~\ref{tab:metrics} are shown in Fig.~\hyperref[fig:metric]{2b} and Fig.~\hyperref[fig:lambda]{3b}, indicating sources localised to within $\unit[1]{deg^2}$ which can be located even faster and more precisely. With \asharp{} (AU) and five Northern Hemisphere observatories, the tight metric is 10\% which corresponds to $\unit[4.2^{+7.6}_{-3.4}]{yr^{-1}}$. The geometric improvement with three or more Northern Hemisphere observatories is greater for the tight metric than the loose metric shown in Fig.~\hyperref[fig:lambda]{3a} although the arithmetic improvement is comparable because the baseline networks' performance is reduced for the tight metric.

\subsection{The next generation (XG)}
Figure~\hyperref[fig:metric]{2c} (Fig.~\hyperref[fig:lambda]{3c}) quantifies the global observation arithmetic (geometric) improvement in the loose metric from adding \cbo{20}{} or \pmo{20}{} in Australia. Both configurations enable significant improvement with a geometric improvement of at least 14 for the global environments of \cbo{40}{; C}; ET; and \cbo{40}{; C} and \cbo{20}{; N}. As expected, the PMO configuration is outperformed here by the CBO configuration which provides a geometric improvement of at least 17 for these environments and an arithmetic improvement of up to 89\% of the sources in range (seen for \cbo{40}{; C} and \cbo{20}{; N}).

The remaining XG global environment of ET and \cbo{40}{; C} observes 90\% of the sources in range by itself. When ET and \cbo{40}{; C} both operate at design sensitivity, therefore, there is little room for improvement in the loose metric using an Australian observatory. For example, adding \cbo{20}{; AU} would improve the metric from $\unit[40^{+71}_{-32}]{yr^{-1}}$ to $\unit[44^{+79}_{-35}]{yr^{-1}}$, a multiplicative factor of only 1.1 because the network then observes 100\% of the sources in range. As discussed in Sec.~\ref{sec:duty_cycle}, however, the duty cycles make it difficult to maximize the observing time with both ET and \cbo{40}{; C} online, and thus the performance of the baseline network is difficult to achieve continuously. This means that an observatory in Australia would significantly improve the amount of time when the global network has two or more XG observatories online. 
If \cbo{40}{; C} [ET] was offline for a major upgrade, then adding \cbo{20}{; AU} to ET [\cbo{40}{; C}] improves the rate of joint observations from $\unit[0.5^{+0.8}_{-0.4}]{yr^{-1}}$ to $\unit[38^{+68}_{-30}]{yr^{-1}}$ [from $\unit[0.2^{+0.3}_{-0.1}]{yr^{-1}}$ to $\unit[19^{+34}_{-15}]{yr^{-1}}$] for a multiplicative factor of 82 [113], as shown in Fig.~\hyperref[fig:metric]{2c} and Fig.~\hyperref[fig:lambda]{3c}. This means that with \cbo{20}{; AU}, ET, and \cbo{40}{; C} it could take less than half a decade to observe $\mathcal{O}(100)$ joint sources even if we do not have all three observatories online at any given time. This is promising for doing cosmology in the XG-era.

Furthermore, the tight metric in Fig.~\hyperref[fig:metric]{2d} and Fig.~\hyperref[fig:lambda]{3d} shows that either Australian CE configuration improves the tight metric more than the loose metric for the baseline network involving both ET and \cbo{40}{; C}. Across all scenarios, the tight metric increases by a geometric factor of at least 2.6, leading to even faster and more precise localisation. Finally, we reemphasise that there is a multitude of science cases, in addition to the one studied in detail in this paper, which will also benefit from an Australian observatory.

\section{Conclusions}
\label{sec:conclusions}
The prospects for multi-messenger gravitational-wave astronomy and cosmology with a future long-baseline global network are promising. We focus on localising binary neutron star mergers to enable electromagnetic follow-up using a gravitational-wave observatory in Australia due to its long baseline to the global network. We define our figure-of-merit as the fraction of binary neutron star mergers available for electromagnetic observation with all channels present, focusing on optical and ultraviolet observations of kilonovae using Rubin. These sources within $\unit[500]{Mpc}$ have a signal-to-noise ratio above 25 and are localised to within $\unit[10]{deg^2}$. 
In the upgraded 2G--era, a $\unit[4]{km}$ arm modified A\# with A+ coatings and a triple-suspension system in Australia significantly improves the number of well-localised sources for networks with 2--4 other observatories by a multiplicative factor of at least 1.4 to beyond 100.
(The improvement is a modest factor of 1.1 when there are five other observatories in the network.)
The gain comes mainly from tighter sky localisation rather than from increasing the network signal-to-noise ratio. The performance using a $\unit[4]{km}$ NEMO is similar since it has comparable broadband sensitivity, however, NEMO is designed for a different science case not addressed here. 
In the XG-era, a broadband-optimised CE $\unit[20]{km}$ in Australia provides a multiplicative improvement of 82--113 when just one other observatory is operating. 
The improvement is marginal when an Australian observatory is added to a network with both ET and a CE in the Northern Hemisphere where 90\% of the sources within $\unit[500]{Mpc}$ are already jointly observed. 
As we discuss in Sec.~\ref{sec:duty_cycle}, however, it is frequently necessary to shut down observatories for maintenance and upgrades, and so an Australian observatory is essential to maintain two simultaneously operating observatories. 
An Australian CE $\unit[20]{km}$ operating alongside ET and a CE in the Northern Hemisphere can be expected to observe $\mathcal{O}(100)$ sources within less than half a decade, even if the three observatories are never all online simultaneously.

We defer to future work a comprehensive Bayesian study of the population within and beyond $\unit[500]{Mpc}$. 
Going beyond $\unit[500]{Mpc}$ would allow us to study multi-messenger astronomy without optical or ultraviolet channels. For example, we could study how the inclination angle affects electromagnetic follow-up since jets and gamma-rays provide a selection effect for face-on/face-off sources~\cite{hotokezaka2019hubble, 10.1093/mnras/stac2613, Taylor_1997, easter2021measure}.
Other avenues of future work include studying network scenarios with a mix of upgraded 2G and XG observatories~\cite{LIGOdovetail}, incorporating early warnings from space-based gravitational-wave observatories, and constructing broadband metrics that combine several science cases for a variety of sources.

This exploratory study, with a focus on multi-messenger astronomy and cosmology, does not address all of the possible science cases or configurations for an Australian observatory.
Other designs focusing on science in the kilohertz regime, e.g., NEMO, would enable studies of extremely dense matter with neutron star mergers, exploring hot and cold equations of state at supranuclear densities. 
Together with the global network, an Australian observatory would enable new science and discoveries.
The improved network sensitivity and increased volume of observations would allow for deeper tests of general relativity in extreme gravity regimes, provide unprecedented insights into the populations and evolution of compact objects, and potentially open new windows onto the dark sector of the Universe to explore fundamental physics.
The promising results presented in this paper encourage future work in this area.

Our code is available online~\cite{repo} and was written using resources from Refs.~\cite{Borhanian_2021, python, ipython, jupyter, numpy, matplotlib, pandas, astropy, bash, slurm}.

\begin{acknowledgements}
The authors wish to thank the following people: B. S. Sathyaprakash, R. X. Adhikari, B. P. Schmidt, S. M. Scott, E. Payne, P. Landry and the providers and support team of OzSTAR. This research is supported by the Australian Research Council Centre of Excellence for Gravitational Wave Discovery (Project No. CE170100004) as well as ARC LE210100002. J.W.G. and this research are supported by an Australian Government Research Training Program (RTP) Scholarship and also partially supported by the United States of America's National Science Foundation under Award No.\ PHY-2011968. S.B. is supported by the Deutsche Forschungsgemeinschaft, DFG, Project MEMI number BE 6301/2-1. P.D.L. is supported by ARC Discovery Project DP220101610. P.D.L. and E.T. are supported by DP230103088. This paper has been assigned LIGO Document Number P2300034.
\end{acknowledgements}

\appendix
\section{NEMO}
\label{app:NEMO}

\begin{figure*}[ht]
    \centering
    \includegraphics[width=\textwidth]{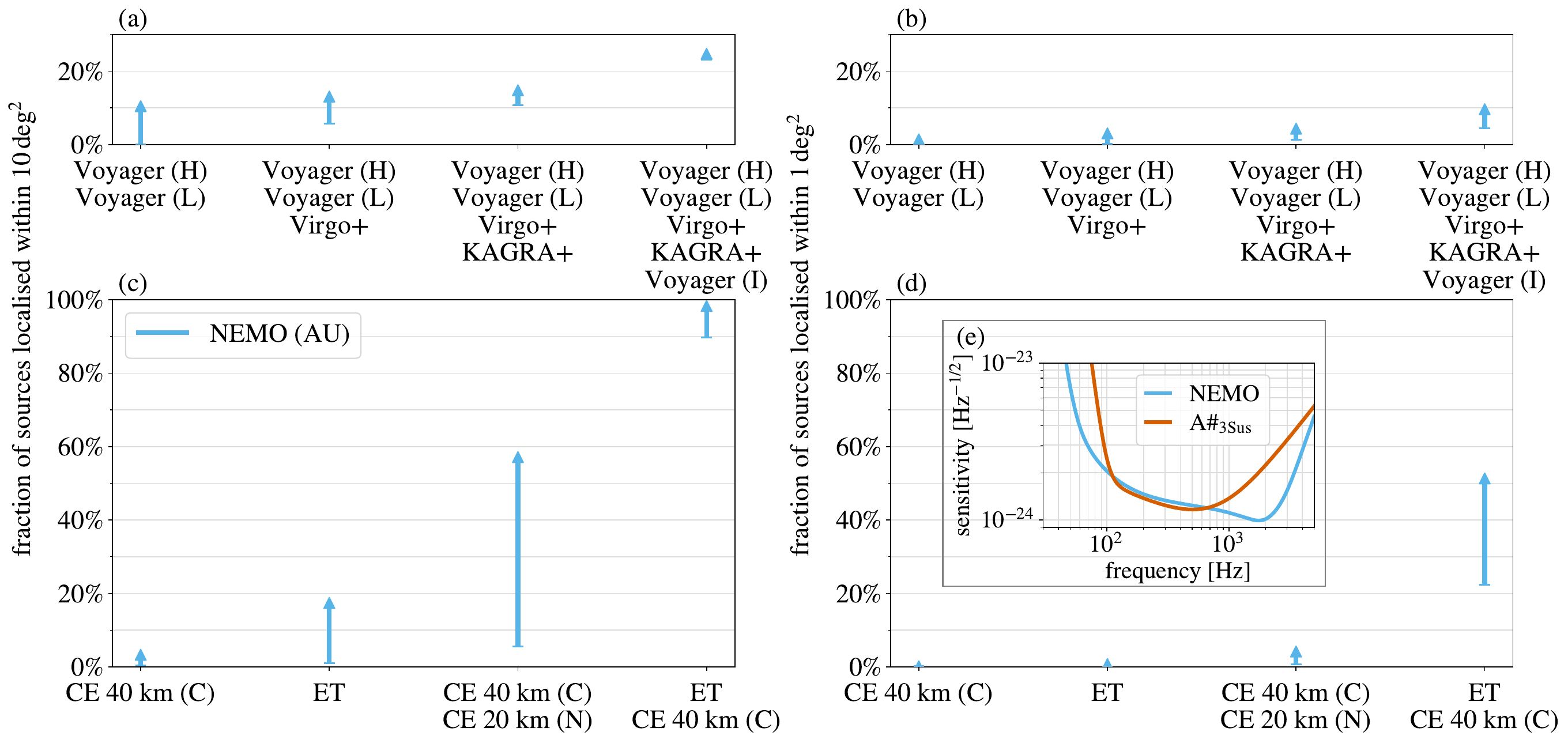}
    \caption{
    (Compare these results to Fig.~\ref{fig:metric}.)
    Arithmetic improvement in the fractional follow-up metrics from adding NEMO in Australia to the baseline global network in Table~\ref{tab:networkScenarios} in the upgraded 2G--era [panels (a) and (b)] and XG-era [panels (c) and (d)].
    The fractional follow-up metrics from Table~\ref{tab:metrics} are the fraction of sources within $\unit[500]{Mpc}$ that have a signal-to-noise ratio greater than 25 and are localised within $\unit[10]{deg^2}$ [panels (a) and (c)] or $\unit[1]{deg^2}$ [panels (b) and (d)].
    Panel (e) shows the strain sensitivity of NEMO and \asharp{} versus frequency.
    }
    \label{fig:NEMO}
\end{figure*}

\begin{figure*}[ht]
    \centering
    \includegraphics[width=\textwidth]{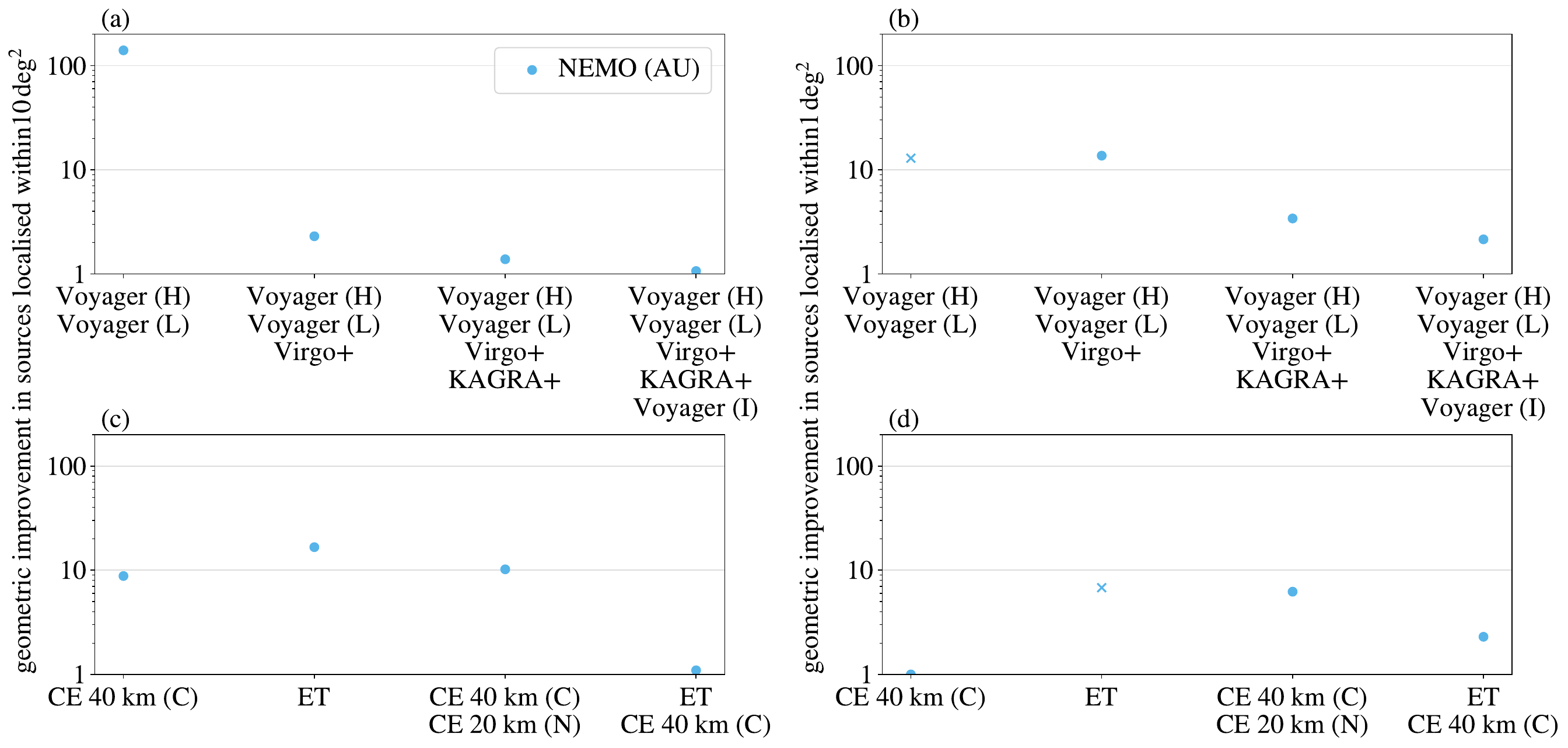}
    \caption{
    (Compare these results to Fig.~\ref{fig:lambda}.)
    Geometric improvement in the follow-up metrics from adding NEMO in Australia to the baseline global network in Table~\ref{tab:networkScenarios} in the upgraded 2G--era [panels (a) and (b)] and XG-era [panels (c) and (d)].
    The geometric improvement in the follow-up metrics from Table~\ref{tab:metrics} is the factor of improvement in the number of sources within $\unit[500]{Mpc}$ that have a signal-to-noise ratio greater than 25 and are localised within $\unit[10]{deg^2}$ [panels (a) and (c)] or $\unit[1]{deg^2}$ [panels (b) and (d)].
    Whenever the baseline network observes zero sources, we provide a lower bound on the geometric improvement indicated by $\times$.
    }
    \label{fig:lambdaNEMO}
\end{figure*}

NEMO is another proposed configuration of the Australian gravitational-wave observatory. It has $\unit[4]{km}$ arms and uses a triple suspension system like \asharp{} but with more advanced internal technology: increased laser power and quantum squeezing, more massive silicon mirrors, and a longer resonant signal-extraction cavity~\cite{NEMO_2020}. As shown in Fig.~\hyperref[fig:NEMO]{4e}, NEMO is more sensitive than \asharp{} in the $\unit[\text{1--4}]{kHz}$ band. NEMO achieves this within the upgraded 2G--era without the significantly higher cost and later start time of the XG observatories. This increased kilohertz sensitivity can enable studying the rich nuclear physics of extreme matter in the intensely hot post-merger phase of binary neutron star coalescences. For example, this would allow us to probe the nuclear equation of state in a high-temperature, high-density regime~\cite{Baiotti2008,Takami2015}. 
We emphasise that we do not consider this post-merger kilohertz science case here.

Although NEMO is not designed for our multi-messenger science case, we still study its performance for completeness. 
For the upgraded 2G--era, the arithmetic improvement in the metrics using NEMO in Fig.~\hyperref[fig:NEMO]{4a--b} compared to \asharp{} in Fig.~\hyperref[fig:metric]{2a--b} are comparable because their broadband sensitivities are similar (see Fig.~\hyperref[fig:NEMO]{4e}). The geometric improvement is also similar between Fig.~\hyperref[fig:lambdaNEMO]{5a--b} and Fig.~\hyperref[fig:lambda]{3a--b}.

For the XG-era, either CE $\unit[20]{km}$ in Australia in Fig.~\hyperref[fig:metric]{2c--d} arithmetically improves the metric more than the $\unit[4]{km}$ arm NEMO in Fig.~\hyperref[fig:NEMO]{4c--d}. The same is true for the geometric improvement shown in Fig.~\hyperref[fig:lambdaNEMO]{5c--d} compared to Fig.~\hyperref[fig:lambda]{3c--d}. \pmo{20}{} has a higher integrated signal-to-noise ratio than NEMO. 
While NEMO's contribution is less than either CE in the XG-era, it would have been operating since the upgraded 2G--era, enabling rich post-merger physics from well before the XG observatories come online.
Future work should further consider the trade-offs between the different science cases and eras of the global network.

\section{Population distribution in redshift}
\label{app:merger_rate}

We simulate a multi-year population of sources out to redshift 0.5, although we only present results for the sources within $\unit[500]{Mpc}$ (redshift $0.1$). For greater resolution in redshift, we sample 250,000 sources distributed uniformly in redshift between 0 and 0.5~\cite{BS2022}. We emphasise that this initial population is not cosmologically accurate. But, after analysing each source, we recover a cosmological distribution.
We use the binary neutron star merger rate $R(z)$ at redshift $z$ from Ref.~\cite{BS2022}. This uses a Madau-Dickinson star formation rate (that does not account for metallicity) and a comoving volume determined by the \texttt{Planck18} cosmology from Ref.~\cite{astropy}.
We divide the redshift range into narrow bins with the endpoints of the bins following a geometric progression. We construct the cosmological distribution by first randomly selecting a bin using the merger rate $R$ as described below and then randomly sampling a source in that bin using a uniform distribution. The probability $p_i$ of drawing the $i$\textsuperscript{th} bin $(z_i, z_i + \Delta_i)$ with width $\Delta_i$ is 
\begin{align}\label{eq:cosmologicalResamplingProbability}
    p_i &= \frac{R_i \Delta_i}{\sum_j R_j \Delta_j} \approx \frac{\int_{z_i}^{z_i+\Delta_i} R(z) \text{d}z}{\int_0^{0.5} R(z) \text{d}z}\\
    \intertext{where the merger rate for each bin is sampled at the geometric mean of the endpoints}
    R_i &= R\left(\sqrt{z_i (z_i + \Delta_i)}\right).
\end{align}
The \emph{multi-year} cosmological population created from this distribution has around 125,000 sources within redshift 0.5 and around 1000 sources within $\unit[500]{Mpc}$.

\section{Computation}
\label{app:computation}
Using the supercomputer OzSTAR at the Swinburne Supercomputing Facility~\cite{ozstar}, we processed the 250,000 uniformly distributed sources. The ``multi-network'' pipeline of \textsc{GWBENCH}~\cite{Borhanian_2021} computes the derivatives of the measured signal for each unique observatory only once. Although calculating these derivatives is the slowest part of the computation, the Fisher information approach remains fast overall compared to Bayesian methods. We took advantage of the multi-network pipeline by grouping similar networks from Table~\ref{tab:networkScenarios} together in each run of 2048 single-core tasks. For this work, we used roughly five runs with each run taking around a day to complete. (Optimising our code could likely lead to a significant speed-up.)
After processing, we cosmologically resampled the results as described in Appendix~\ref{app:merger_rate}.

\section{Rejections}
\label{app:rejections}
\begin{figure*}[ht]
    \centering
    \includegraphics[width=\textwidth]{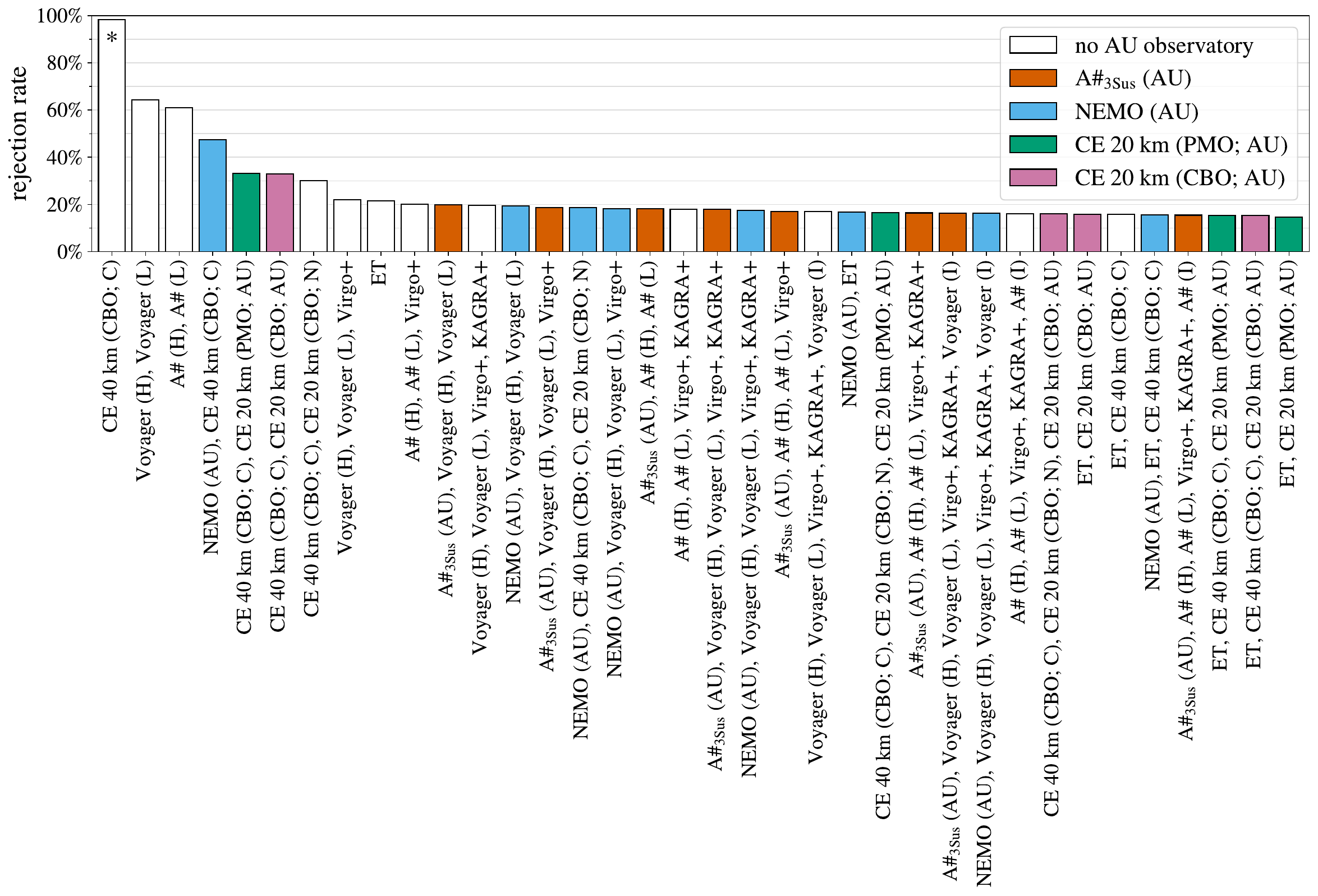}
    \caption{Rejection rate (i.e., the fraction of sources in range that are numerically ill-conditioned) for each upgraded 2G--era and XG-era network shown in Table~\ref{tab:networkScenarios} in descending order. We also include the other 2G-era networks that use A\# instead of Voyager for the LIGOs (although they are omitted from the observation rate results) as well as the networks involving NEMO from Appendix~\ref{app:NEMO}. The single-observatory network indicated by $\ast$ is excluded from the unification of the rejections.} 
    \label{fig:rejectionRate}
\end{figure*}

In Section~\ref{sec:fisher}, we invert the Fisher information matrix to calculate the Cram\'er-Rao bound~\cite{cramer1946mathematical,rao}. 
This matrix can be numerically ill-conditioned and inaccurate to invert~\cite{vallisneri2008use,rodriguez2013inadequacies}. 
In such cases, we reject the source from the analysis, not including it in the cosmological resampling or the results. 
Fig.~\ref{fig:rejectionRate} shows the rejection rate, i.e., the percentage of sources in range that are ill-conditioned, for each network scenario.
For the single-observatory network \cbo{40}{; C}, the estimates of the sky position are degenerate and the rejection rate is $98\%$. (ET comprises an equilateral triangle of three identical observatories at the same site such that the overall effective sensitivity shown in Fig.~\ref{fig:sensitivityCurves} is a factor of $\frac{2}{3}=\left(\sqrt3\sin\left(\frac{\pi}{3}\right)\right)^{-1}$ more sensitive than each individual observatory~\cite{EinsteinTelescope}.) For the multiple-observatory networks, the rejection rate is 15--64\% (networks with more observatories tend to have a lower rejection rate). 
These rates are for the population uniformly distributed in redshift because we reject the sources before the cosmological resampling (see Appendix~\ref{app:merger_rate}). We assume that the rejection rate is the same for the cosmological population and defer to future work eliminating the rejections by using Bayesian analysis instead. 
Switching to a Bayesian framework would also allow us to study sources with lower signal-to-noise ratios.
One alternative to eliminate the rejections is to use arbitrary-precision floating-point arithmetic~\cite{mpmath,gwbenchv7}.

For Fisher information methods, rejections are a perennial issue among other challenges~\cite{vallisneri2008use,rodriguez2013inadequacies}.
To ensure a fair comparison, we only study sources that are unrejected for all multiple-observatory networks. This ``unified'' rejection rate is 68\%. This guarantees that the cosmological population is the same for each multiple-observatory network.
(We include the other 2G-era and NEMO networks but exclude the single-observatory network.)
Choosing a smaller set of networks among which to unify the rejections would give different results. But, it is necessary that we are able to compare any two multiple-observatory networks. 
Unifying the rejections is also necessary to make the geometric improvement invariant of the rejection rate.

\begin{figure}[ht]
    \centering
    \includegraphics[width=\columnwidth]{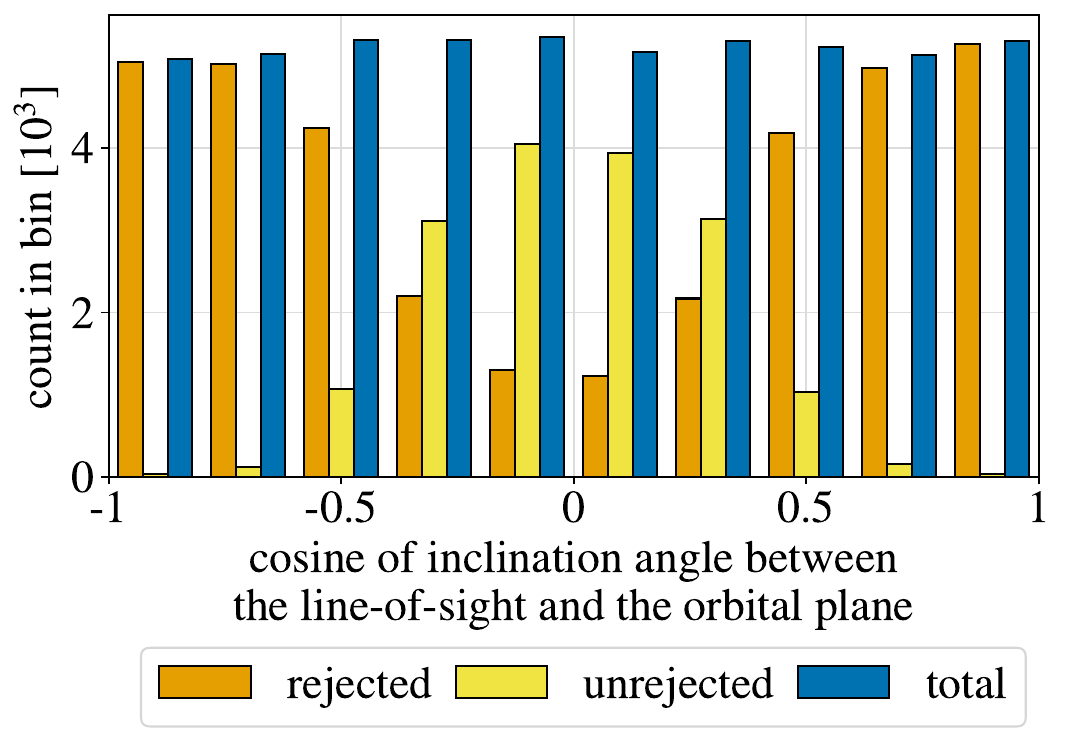}
    \caption{Histogram of the number of uniformly rejected and unrejected sources in range versus the inclination angle. The initial population (labelled as ``total'') is sampled from a uniform distribution in the cosine of the inclination angle. 
    There is a bias towards rejecting nearly edge-on sources, corresponding to an inclination angle of close to $0$ or $\pi$.}
    \label{fig:rejectionProfileInclination}
\end{figure}

We profile what kinds of sources are rejected.
The dominant effect is the rejection of nearly edge-on sources shown in Fig.~\ref{fig:rejectionProfileInclination}. 
Nearly edge-on sources have a lower signal-to-noise ratio than their counterparts with a more preferred orientation. Our metrics, therefore, may be overestimated in some cases. The ``null-case'' scenario is that --- at the population level --- the rejected sources behave the same as the unrejected sources. This would mean that the metrics are unaffected. The worst-case scenario, however, is that none of the rejected sources can be jointly observed. This would reduce the fractional metrics by a multiplicative factor of 0.32.

For example, for \asharp{} (AU) added to the network of Voyager (H), Voyager (L), and Virgo+ (see Fig.~\ref{fig:metric}), the null (worst) case scenario is $\unit[5.6^{+10}_{-4.5}]{yr^{-1}}$ ($\unit[1.8^{+3.2}_{-1.4}]{yr^{-1}}$). The combined uncertainty interval accounting for both scenarios would be from $\unit[0.4]{yr^{-1}}$ to $\unit[16]{yr^{-1}}$.
Since we presently have insufficient evidence to reject the null hypothesis, however, we use the null-case scenario in the results.
We defer determining the true rejection scenario or eliminating the rejections to future work.

\end{document}